\documentclass[12pt,reqno,a4paper]{amsart}
\title[Regression]{On parameter estimation in the physics lab based on inverting a slope regression coefficient}
\author{W. Jacquet, E. Nyssen, \& J. Sijbers}

\usepackage{graphics}
\usepackage{graphicx}
\usepackage{amssymb}
\usepackage{amsfonts}
\usepackage{epsfig}
\usepackage{color}
\usepackage{epstopdf}

\def\epsilon{\varepsilon}
\def\phi{\varphi}
\def\theta{\vartheta}
\def\rho{\varrho}

\def\beq#1{
\begin{equation}\label{#1} {\tt\hspace*{-6em{[#1]~~~}}}
}
\def\eeq{\end{equation}}

\def\beq#1{\begin{equation}\label{#1}}
\def\eeq{\end{equation}}

\def\phi{\varphi}
\def\theta{\vartheta}

\usepackage[normalem]{ulem}

\begin{document}

\begin{abstract}
Measurement uncertainty is a non trivial aspect of the laboratory component of most undergraduate physics courses. 
Confusion about the application of statistical tools calls for the elaboration of guidelines and the elimination of inconsistencies were possible. 
Linear regression is one of the fundamental statistical tool often used in a first year physics laboratory setting.
In what follows we present an argument that leads to an unambiguous choice of (a) variable(s) to be used as predictor(s) and variable to be predicted.  
\vspace{2em}

\end{abstract}

\maketitle

\section{Introduction}
\setcounter{equation}{0}

A great number of physical laws that freshmen are confronted with in the physics lab are linear in nature or can be linearized.
The parameters of these relations are physical constants or characteristics of the system under study.
Most often the aim of the lab is to estimate these constants and if not,
the estimation of these constants is an intermediate step.
Therefore, linear regression is a nearly inevitable statistical tool. 
It enables to estimate the constants and to evaluate the validity of the linear relation.
A non-constant linear function has two formulations, it can be inversed and its inverse is linear with as
slope the inverse of the slope of the original function;
if $ y = a x + b$ with $a \ne 0$ then $ x = (1/a) y - b/a$.
It is common in physics labs to use the formulation of a linear relations that emerges naturally as a basis for the regression inconsiderate whether $a$ or $1/a$ is of interest.
Based on theoretical considerations or habit the consequence of an action is place on the left side as a function of the cause(s) on the right side.
If the formulation presented to the students is $ y = a x + b$ and $1/a$ has to be estimated a great number of students first estimates $a$ and subsequently determines the inverse as an estimate for $1/a$. 
The assessment of the uncertainty is based on the uncertainty of $a$ and standard error propagation.
We will refer to this approach as the ``inverse slope'' strategy in contrast with the ``direct'' estimation strategy. 

Both strategies of estimation are compliant with the international standard ``ISO Guide to the Expression of Uncertainty in Measurement'' (GUM) that provides a basis for the assessment of uncertainty.\cite{GUM1995} The GUM however does not treat regression analysis. 
The very accessible introduction to the treatment of experimental data by Young assumes that the measurement uncertainty of the argument in the linear relation can be neglected.\cite{Young1962} 
Also more recent treatises of uncertainty in physical measurements attribute the error to one of the variables only -- see e.g.\ \cite{Squires1985}.
Rabinovich explicitly mentions the strategy of inversion in a calibration setting.\cite{Rabinovich2005}
It must be clear that if standard linear regression requires the measurement uncertainty to be insignificant for all but one and only one variable, 
the variable to be predicted, the advisable strategy starts with using the variable with the ``least'' measurement uncertainty as argument in the linear relation. 
However, measurement uncertainty is dependent on the unit of measurement.
Therefore, if different measurement units are used for the different variables involved, comparison of uncertainties as such is not possible. 
The obvious method to eliminate the measurement units is the use of relative errors, yet these errors are not shift invariant. 
Moreover, whenever the observed values are near zero the relative error will tend to be high. 
Therefore, relative errors are not a good criterion to determine which variable(s) to use as an argument.
When looking at standard multivariate statistical literature the error in the linear regression model is not specifically attributed to one or the other variable -- see e.g.\
\cite{Bhattacharyya1977}.
In Foranasini uncertainty is attributed to both variables involved and shifted to one side of the equation.\cite{Fornasini2008}
Gill pleads to use the regression in a direction opposite to the natural causality when prediction is inverse to the causation.\cite{Gill1987}

In what follows we will follow the strategy of Foranasini and stress the importance of choosing the appropriate formulation when estimating constants describing a system. 
In contrast to Gill our paper focuses on the proximity of zero as a disturbing factor in the inversion. 
Also using elementary mathematical tools inversion bias is revealed.  
The problem will first be approached from a theoretical point of view and illustrated through the simulation of a classical experiment for estimating the
gravitational constant.

\section{Theoretical considerations}
\setcounter{equation}{0}
\subsection{Definition of the problem}
Let $y = a x + b$ with $a \ne 0$, $a$ and $b$ are unknown, and $1/a$ is of interest.
\subsection{Regression model}
Either the values of the variables $y$ and $x$ are measured simultaneously, either one is set and the other measured.
Errors will occur when a variable is measured and also if a variable is set, 
the exact effective value will only be known up to an error:
\begin{equation}
\begin{array}{ccc}
X = x + \epsilon_x & \mbox{ and } & E[\epsilon_x]=0 \mbox{ , } \\
\end{array}
\label{MF1}
\end{equation}
\begin{equation}
\begin{array}{ccc}
Y = y + \epsilon_y & \mbox{ and } & E[\epsilon_y]=0 \mbox{ , } \\
\end{array}
\label{MF2}
\end{equation}
where $\epsilon_x$ and $\epsilon_y$ are independent.
After substitution of $x$ and $y$ into the linear relation the following regression models are obtained:
\begin{equation}
\begin{array}{ccc}
Y = a X + b     + [\epsilon_y - a \epsilon_x  ] & \mbox{ let } & \epsilon_Y=[a \epsilon_x -\epsilon_y]   \mbox{ (notation Y/X) , } \\
\end{array}
\label{RM1}
\end{equation}
\begin{equation}
\begin{array}{ccc}
X = (1/a) Y - b/a + [  \epsilon_x -\epsilon_y/a] & \mbox{ let } & \epsilon_X=[  \epsilon_x -\epsilon_y/a] \mbox{  (notation X/Y) . } \\
\end{array}
\label{RM2}
\end{equation}
Notice that $E[\epsilon_X] = E[\epsilon_Y]=0$ and if  $\epsilon_x$ and $\epsilon_y$ are normally distributed then so are $\epsilon_X$ and $\epsilon_Y$.
Standard OLS linear regression requires a linear relation between the variable predicted and the variable measured up to a random variable. 
Equations \ref{RM1} and \ref{RM2} however show that in both formulations the variable used as predictor is itself a random variable correlated with the ``error'' terms $\epsilon_X$ and $\epsilon_Y$. 
The results in a bias towards zero of the coefficient in the linear relation. 
This phenomena is well known and generally referred to as ``error in variables'' problem -- see e.g. \cite{Fuller1987}.
In order to use ordinary least squares regression the measurement error of the variable used as argument in the linear relation has to be ``sufficiently small''. 
An elaborate treatment and approximation of error made can be found in \cite{Davies1975}. 
Most often it is not appropriate to introduce these techniques at a general physics course and repeated measurements are used to reduce the measurement error. 
In what follows we will suppose that the error on the arguments are sufficiently reduced through repeated measurement.
If the specific laboratory setting does not allow for precise measurements other more appropriate statistical techniques involving generalized linear models should be used. 

\subsection{Bias}
If a random variable $u$ has a continuous density function $f$ which is
strictly positive at zero, it can be shown that  $E[1/u]=\int_{-\infty}^{+\infty}(1/u) f(u) du$ does not exist.
Follows that the bias of $1/\hat{a}$ , with $\hat{a}$ the regression estimator of $a$, does not exist.
Although measurement errors are most often assumed to be normally distributed sticking to the former result is not satisfactory.
Errors can not become arbitrarily large in practice and when a value is unrealistic the measurement is discarded and repeated.
Therefore, let us assume that $\hat{a}$ is symmetrically distributed about its expectation $a$ and
has only non zero density in a finite interval $[a-d,a+d]$ that does not contain zero.
Note that as a consequence, the variance $\sigma^2$ exists.

Consider the second order Taylor expansion of $1/u$ at  $u=a$:
\begin{equation}
\begin{array}{ccc}
\frac{1}{u} =  \frac{1}{a} - \frac{1}{a^2} (u-a) + \frac{1}{\xi^3} (u-a)^{2} & \mbox{ with } & \xi \in ]u,a[ \\
\end{array}
\label{G2}
\end{equation}
\begin{equation}
\begin{array}{ccc}
\frac{1}{u} -  \frac{1}{a} + \frac{1}{a^2} (u-a) = \frac{1}{\xi^3} (u-a)^{2} & \mbox{ with } & \xi \in ]u,a[ \\
\end{array}
\label{G3}
\end{equation}
Follows:
\begin{equation}
\begin{array}{ccc}
\frac{1}{u} -  \frac{1}{a} + \frac{1}{a^2} (u-a) > \frac{1}{(a+d)^3} (u-a)^{2} & \mbox{ for all } & u \in $]a-d,a+d[$ \\
\end{array}
\label{G4}
\end{equation}
The expectation of the left and the the right side of the inequality results in:
\begin{equation}
\begin{array}{c}
E[\frac{1}{u}] -  \frac{1}{a}  > \frac{1}{(a+d)^3} \sigma^2  \\
\end{array}
\label{G5}
\end{equation}

A large variance for $u$ will therefore result in a large bias.
Bias does not tell the whole story. 
Most often we are interested in a confidence interval.
An interval estimate of $a$ can be transformed into an interval estimate of $1/a$ without loss of accuracy.
An interval that contains $a$ and does not contain zero is mapped to an interval that contains $1/a$ through $1/u$.
A process that generates  an interval that will contain $a$ with probability $p=95\%$, generates after transformation $1/u$ an interval about $1/a$ with a $p=95\%$ probability.
Such an interval is not presented as ``estimated value $\pm$ error'', expected from a physics experiment.
When the classical error propagation is used to transform a confidence interval about $\hat{a}$ into a confidence interval about $1/\hat{a}$,
two observations have to be made: 
$1/\hat{a}$ is not an unbiased estimator of $1/a$ and for ``small'' values of $u$ the function $1/u$ is not well approximated by its tangent, the approximation being basis of the classical error propagation.

\section{Estimation of the gravitational constant}
\setcounter{equation}{0}

A body in the neighborhood of the earth experiences a nearly constant gravitational force and its acceleration is constant provided that all other forces are neglected.
This constant acceleration parameter is called the gravitational constant $g$.
Consider an object that falls from a height $h$, starting at rest, and the time $t$ it needs to cross this distance.
After a small manipulation of Newton's law, 
the following relation between the height $h$ an object starting at rest falls and the time $t$ needed to fall this height is obtained:
    \begin{equation}\label{Reg_h_t2}
    h = \frac{1}{2} \, g \, t^2  \mbox{ . }
    \end{equation}
When the height $h$ and the square of the fall time $t^2$ are obtained for different values for the height it is possible to estimate the gravitational constant through linear regression.
The first approach consists of a linear regression $y = a\,x$ with $h$ used as $y$ variable, 
$t^2$ as $x$ variable, and $g=2a$. 
Equation \ref{Reg_h_t2} can also be rephrased:
    \begin{equation}\label{Reg_t2_h}
    t^2   = 2 \, \frac{1}{g} \,  h \mbox{ . }
    \end{equation}
This leads to a second approach to the estimation of $g$.
First $2/g$ is estimated through linear regression $y=a\,x$ with $t^2$ used as $y$ variable and $h$ as $x$ variable.
The estimate of $g$ is obtained through inversion and doubling of the estimate of $a$: $g = 2/a$.

\subsection{Numerical simulation experiment}
\begin{figure}
\begin{center}
\begin{tabular}{c}
\includegraphics*[width=4cm]{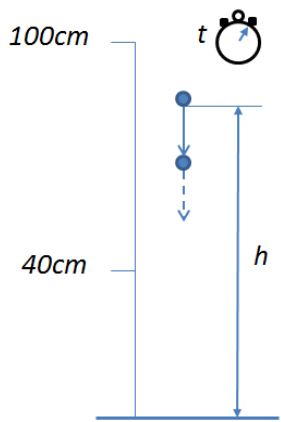}
\end{tabular}
\end{center}
\caption{\small\it A steel bullet is dropped from a height $h$ and the fall time $t$ is measured. \label{Fig:Drop1}}
\end{figure}

All experiments were conducted using  MatLab R2010a (The Mathworks, Inc., Natick, Massachusetts).
A steel bullet is dropped from 10 uniform randomly chosen heights between $0.4m$ en $1m$ in order to estimate the gravitational constant --see Fig.\ \ref{Fig:Drop1}.
To simulate this experiment, 10 random numbers between $0.4$ and $1$ are generated.
For each number a value for the fall time $t$ is calculated using equation \ref{Reg_t2_h} with $g=9.81 m/s^2$.
Height and time are both measured with a measurement error.
In what follows, the measurement error of height and time are assumed to be normally distributed with mean zero and standard deviation $\sigma_h$ and $\sigma_t$ respectively.
The generated values for height and time are disturbed by adding values generated from a normal distribution with zero mean and standard deviation $\sigma_h$ and $\sigma_t$ respectively.
The standard deviation for the time $t$ is chosen to be $0.0001s$.
This relatively small error models a very accurate time measurement.
The former experiment is repeated 1000 times for different values for $\sigma_h$.
For a standard deviation $\sigma_h=0.01 m$, ten repetitions of 1000 experiments were performed.

\noindent {\bf Results.}
The results of the experiments are presented in tables \ref{Tab:1}, \ref{Tab:2} and \ref{Tab:3}.
Although for all criteria the observed standard deviations are comparable for 
both estimation methods, the standard deviation of the difference between the gravitational constant and the estimated gravitational constant is systematically smaller when estimated based on the regression $h$ given $t^2$ ($h/t^2$).
The error decreases with decreasing error in height for both methods, but the error obtained through the regression $t^2$ given $h$ ($t^2/h$) is systematically higher than the error obtained through $h/t^2$.
For all chosen values for the standard deviation $\sigma_h$ the mean difference of the gravitational constant and the estimated gravitational constant is positive, although never statistically significant ($p<.05$).
When repeating the series of 1000 experiments 10 times for $\sigma_h = 0.01m$ only one of the mean differences was negative --see table \ref{Tab:2}.
The estimation of the gravitational constant based on ($h/t^2$) is positively biased (Sign test N=10, p=0.21).

\begin{table}
\begin{center}
\begin{tabular}{ccccccc}
height $h$ & \multicolumn{2}{c}{estimate $g$} 		 & \multicolumn{2}{c}{error} 		 & \multicolumn{2}{c}{difference} 		\\
 std	 & mean	 & std	 & mean	 & std	 & mean	 & std	\\
 ($m$)	 & ($m/s^2$)	 & 	 & ($m/s^2$)	 & 	 & ($m/s^2$)	 & 	\\
  $ 0,1 $ 	 & $ 10,116 $ 	 & $ 0,573 $ 	 & $ 0,640 $ 	 & $ 0,162 $ 	 & $ 0,306 $ 	 & $ 0,573 $ 	\\
  $ 0,05 $ 	 & $ 9,892 $ 	 & $ 0,303 $ 	 & $ 0,333 $ 	 & $ 0,082 $ 	 & $ 0,082 $ 	 & $ 0,303 $ 	\\
  $ 0,03 $ 	 & $ 9,877 $ 	 & $ 0,282 $ 	 & $ 0,323 $ 	 & $ 0,081 $ 	 & $ 0,067 $ 	 & $ 0,282 $ 	\\
  $ 0,01 $ 	 & $ 9,815 $ 	 & $ 0,054 $ 	 & $ 0,058 $ 	 & $ 0,014 $ 	 & $ 0,005 $ 	 & $ 0,054 $ 	\\
  $ 0,005 $ 	 & $ 9,811 $ 	 & $ 0,025 $ 	 & $ 0,028 $ 	 & $ 0,007 $ 	 & $ 0,001 $ 	 & $ 0,025 $ 	\\
  $ 0,001 $ 	 & $ 9,810 $ 	 & $ 0,005 $ 	 & $ 0,005 $ 	 & $ 0,001 $ 	 & $ 0,000 $ 	 & $ 0,005 $ 	\\
\end{tabular}
\end{center}
%EEE\label{Tab:1}
\caption{Estimation of $g$ by regressing $t^2$ on $h$\label{Tab:1}}
\end{table}

\begin{table}
\begin{center}
\begin{tabular}{ccccccc}
height $h$ & \multicolumn{2}{c}{estimate $g$} 		 & \multicolumn{2}{c}{error} 		 & \multicolumn{2}{c}{difference} 		\\
 std	 & mean	 & std	 & mean	 & std	 & mean	 & std	\\
	 & ($m/s^2$)	 & 	 & ($m/s^2$)	 & 	 & ($m/s^2$)	 & 	\\
  $ 0,1 $ 	 & $ 9,824 $ 	 & $ 0,571 $ 	 & $ 0,620 $ 	 & $ 0,148 $ 	 & $ 0,014 $ 	 & $ 0,571 $ 	\\
  $ 0,05 $ 	 & $ 9,809 $ 	 & $ 0,301 $ 	 & $ 0,330 $ 	 & $ 0,080 $ 	 & $ -0,001 $ 	 & $ 0,301 $ 	\\
  $ 0,03 $ 	 & $ 9,799 $ 	 & $ 0,280 $ 	 & $ 0,320 $ 	 & $ 0,079 $ 	 & $ -0,011 $ 	 & $ 0,280 $ 	\\
  $ 0,01 $ 	 & $ 9,812 $ 	 & $ 0,054 $ 	 & $ 0,058 $ 	 & $ 0,014 $ 	 & $ 0,002 $ 	 & $ 0,054 $ 	\\
  $ 0,005 $ 	 & $ 9,810 $ 	 & $ 0,025 $ 	 & $ 0,028 $ 	 & $ 0,007 $ 	 & $ 0,000 $ 	 & $ 0,025 $ 	\\
  $ 0,001 $ 	 & $ 9,810 $ 	 & $ 0,005 $ 	 & $ 0,005 $ 	 & $ 0,001 $ 	 & $ 0,000 $ 	 & $ 0,005 $ 	\\
\end{tabular}
\end{center}
\caption{Estimation of $g$ by regressing $h$ on $t^2$.\label{Tab:2}}
\end{table}

\begin{table}
\begin{center}
\begin{tabular}{cccc}
\multicolumn{2}{c}{estimate $g$} 		 & \multicolumn{2}{c}{error} 		\\
mean 		  & std		   & mean 	   & std		\\
 ($m/s^2$)  	  & 	 	   & ($m/s^2$)	   & 	 	\\
   $9,815$ 	   & $0,054$ 	   & $0,005$ 	   & $0,054$ 	\\
   $9,811$ 	   & $0,056$ 	   & $0,001$ 	   & $0,056$ 	\\
   $9,812$ 	   & $0,053$ 	   & $0,002$ 	   & $0,053$ 	\\
   $9,814$ 	   & $0,054$ 	   & $0,004$ 	   & $0,054$ 	\\
   $9,815$ 	   & $0,056$ 	   & $0,005$ 	   & $0,056$ 	\\
   $9,809$ 	   & $0,055$ 	   & $-0,001$ 	   & $0,055$ 	\\
   $9,811$ 	   & $0,049$ 	   & $0,001$ 	   & $0,049$ 	\\
   $9,814$ 	   & $0,058$ 	   & $0,004$ 	   & $0,058$ 	\\
   $9,814$ 	   & $0,048$ 	   & $0,004$ 	   & $0,048$ 	\\
   $9,812$ 	   & $0,064$ 	   & $0,002$ 	   & $0,064$ 	\\
\end{tabular}
\end{center}
\caption{Estimation of $g$ by regressing $t^2$ on $h$ for 10 repetitions with $\sigma_h = 0.01m$. \label{Tab:3}}
\end{table}

\section{Discussion and conclusions}
\setcounter{equation}{0}

Theory and our numerical experiment show that the choice of formulation can be important when estimating through linear regression.
We proved and demonstrated experimentally that estimation followed by inversion introduces bias and an increased reported error.
The general rule should be that if the value of a variable or constant can be estimated avoiding inversions one should do so since when the original estimate is unbiased the inverse estimate will be biased.
Although differences between methods are not excessive there are no advantages in using a regression followed by an inversion in comparison to the direct estimation.
The combination of regression and inversion scores systematically worse than direct estimation.
The laboratory experiment for deriving the gravitational constant
is an example of a very common exercise for students studying physics.
The most straightforward way of conceiving such experiment consists of
measuring the time of fall of a bullet for a given set of height values.
This suggests time as natural ``dependent variable'' of height and inspires
an approach where $ a = 2/g $ is estimated (according to Eq.\ \ref{Reg_t2_h}).
We have shown that this approach introduces a bias in the estimated
value $ \hat{g} $.
The attentive reader may have noticed that for realistic values of $ \sigma_h $ (e.g.\ 0.001m) this bias can be neglected for the
gravity constant experiment.
However, the point we want to make here is that the students should be warned about the existence of this bias and should be made aware of
the fact that this bias -- which is misleadingly small in an experiment
like the one aiming at estimating the gravity constant -- may be important in
other types of experiments where one equally tends to calculate
$ 1/\hat{a} $ as estimate of a parameter.

\bibliographystyle{plain}
\bibliography{Error}

\newpage

\noindent Wolfgang Jacquet\\
Vision Lab,
Department of Physics, \\
University of Antwerp Universiteitsplein 1, \\
B-2610 Wilrijk, \\
Belgium. \\
E-mail: wolfgang.jacquet@ua.ac.be, \\
Phone: +32 (0)3 265.24.77 \\
Fax: 	+32 (0) 3 265 22 45 \\
Office: D.N.1.18 Campus Drie Eiken \\

\noindent E. Nyssen,\\
Vrije Universiteit Brussel, \\
Department of Electronics and Informatics - ETRO,\\
Pleinlaan 2,\\
B-1050 Brussels,\\
Belgium.

\noindent J. Sijbers, \\
Vision Lab,
Department of Physics, \\
University of Antwerp Universiteitsplein 1, \\
B-2610 Wilrijk, \\
Belgium.

\end{document}